\documentclass[preprint]{elsarticle}

\usepackage{amsmath}
\usepackage{graphicx}
\usepackage{hyperref}
\usepackage{indentfirst}
\usepackage{amssymb}

\newcommand{\vev}[1]{\langle #1 \rangle}
\newcommand{\ket}[1]{\vert #1 \rangle}

\begin{document}

\title{Graceful exit via monopoles in a theory with O'Raifeartaigh type supersymmetry breaking}

 \author{Brijesh Kumar}
 \ead{brijesh@phy.iitb.ac.in}
 \author{Urjit A. Yajnik}
 \ead{yajnik@phy.iitb.ac.in}
 \address{Department of Physics, Indian Institute of Technology 
Bombay, Mumbai - 400076, India \\ and \\ Indian Institute of Technology Gandhinagar, 
Ahmedabad - 382424, India}


\begin{abstract}
We study the stability of supersymmetry breaking metastable vacua and
supersymmetric vacua in the presence of solitons. The metastable vacua
of supersymmetric QCD and those found elsewhere such as in models based 
on the SU(5) grand unified group support the existence of topological solitons. 
The vacua containing such topological defects
can become unstable against decay into lower energy configurations. 
We show for a specific model that a finite region of the available parameter space
of couplings becomes disallowed due to the presence of 
monopoles. In a manner similar to previous studies based on cosmic strings, 
it is shown that soliton solutions arising in supersymmetric theories can 
put constraints on the range of allowed values of the couplings arising 
in the theories. Implications for cosmology are discussed.

\end{abstract}

\maketitle

\section{Introduction}

Dynamical supersymmetry breaking in metastable vacua is an effective
means of breaking supersymmetry which can be accommodated in various
classes of models such as $\mathcal{N} = 1$ supersymmetric $SU(N_c)$
QCD with $N_f$ massive fundamental flavors \cite{Intriligator:2006dd,Intriligator:2007py}. 
Models based on this method can have several supersymmetry breaking false
vacua which have a finite lifetime based on the quantum tunneling rate to
the true vacuum \cite{Coleman}. Metastable vacua which break supersymmetry
also occur in many other models of supersymmetry breaking and mediation
\cite{Dine:1995ag,Dine:1994vc,Luty:1998vr,Banks:2005df}.
Issues relating to the phenomenological implementation of these ideas have
been discussed in \cite{Dine:2007dz,Banks:2006ma,
Banks:2005df,Intriligator:2007py,Intriligator:2006dd, Dine:2006gm,
Shih:2007av,Dine:2006xt,Essig:2007kh,Aharony:2006my,
Dasgupta:1996pz,Craig:2006kx,Abel:2006cr} and cosmological
applications appear in \cite{Ferrantelli:2009zv, Fischler:2006xh, Endo:2007sw}.
A necessary condition for the viability
of such models is that the lifetime of the metastable vacua is much
larger than the age of the universe. 

In a recent paper \cite{Kumar:2008jb}, we had emphasized that it is not sufficient
to study the stability of supersymmetry breaking vacua in terms of their
translation invariant form alone. In general, topological defects can form
in the evolution of the early universe \cite{Kibble:1980mv}.
For example, the models of supersymmetry breaking described in
\cite{Intriligator:2006dd,Intriligator:2007py} break SUSY in metastable vacua 
which can contain cosmic strings \cite{Eto:2006yv}. Such defects
have important consequences with regards to stability issues. The core
of a cosmic string or monopole can give rise to a ``seeding'' of the
true vacuum and render it unstable against decay into a lower energy
configuration \cite{Yajnik:1986wq, Steinhardt, Steinhardt:1981ec}.
This is true even if the translation
invariant vacuum is sufficiently stable against quantum mechanical tunneling
to the true vacuum. Such a process has been
studied in the context of phase transitions in Grand Unified
Theories in \cite{Yajnik:1986tg}. Instabilities of domain walls in theories with
compact extra dimensions are discussed in \cite{Aguirre:2009tp}.

In \cite{Kumar:2008jb} we dealt with the messenger sector of a model in which supersymmetry
breaking occurs according to the gauge mediated supersymmetry breaking scenario. The
seeding resulted from the presence of cosmic string solutions. As an extension of this
work, we study the seeding mechanism in the presence of monopoles, and this is done for 
the case of direct supersymmetry breaking rather than having supesymmetry breaking being 
communicated from a hidden sector. In particular,
we have chosen to study a non-abelian model in which both the gauge and supersymmetry breaking
occur simultaneously in a O'Raifeartaigh type model through vacuum expectation values of
appropriate Higgs scalars. After $SU(5)$ breaking, there are no supersymmetric
vacua in this model. There are however multiple supersymmetry breaking
vacua, some of which are metastable. It turns out that
some of these metastable and spatially homogeneous vacuum states become disallowed
in the presence of monopoles for a large range of couplings occurring in
the model. In addition to the instability 
which can arise in supersymmetry breaking vacua, we show in the present work
that this effect can also take place for two supersymmetric vacua.
It is possible to study the stability
numerically and also semi-analytically. Both of these approaches
will be dealt with in this paper.

The classical instabilities of solitons discussed here can have a significant impact
on the evolution of the early universe. In the model we deal with, there is a metastable
supersymmetry breaking vacuum which is undesirable from the point of view of phenomenology. It
is possible for some causally connected regions of the universe to get trapped in this 
false vacuum during the early stages of
its evolution. If no monopoles are present, 
the transition to the true vacuum can only take place via quantum tunneling. 
This can result in a potentially inhomogeneous universe with the first-order phase transition never
being completed \cite{Guth:1982pn, Hawking:1982ga}.
As we will show, the presence of monopoles in such metastable vacua can result in classical
instabilities of such configurations and hence a graceful exit from a potentially inhomogeneous
expansion.

The next section \ref{classinst} discusses the classical instabilities of solitons in general. 
The following section \ref{modelandansatz} then briefly describes the model and its vacua,
while section \ref{eom} deals with
the monopole ansatz and equations of motion. Sections \ref{nonsusyg}
and \ref{stabilitysusy}
discuss the stability of the vacua containing monopoles from a
numerical and semi-analytic approach. Implications for cosmology 
are presented in section \ref{impl} followed by concluding remarks in 
section \ref{conc}.

\section{Classical Instabilities of Monopoles and Vortices}
\label{classinst}

We begin with brief overview of how vortex and monopole solutions can become unstable
classically under certain conditions. The dissociation of $SU(5)$ monopoles which
will be discussed in this paper is similar to the monopole dissociation studied
in \cite{Steinhardt}. A general study of this phenomenon which applies for both
vortices and monopoles is discussed in \cite{Steinhardt:1981ec}. 

\begin{figure}[!htp]
\begin{center}
\includegraphics[width=0.8\textwidth]{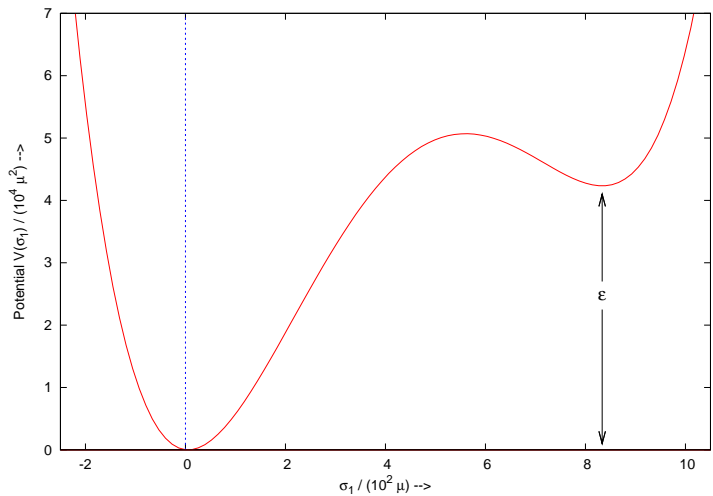}
\caption{An example of the potential to be studied in this paper. The value of $\epsilon$ is
given by the difference in energy densities of the false and true vacuum. The term $\mu$ is
a mass scale which will appear in the model discussed here.}
\label{potcur}
\end{center}
\end{figure}

Let us consider the scalar potential for a field $\sigma_1$ as shown in figure 
\ref{potcur}. This potential is an example of the type of potential we will encounter in 
this study. It has a global minimum of nearly vanishing energy for which the VEV of 
$\sigma_1$ is close to zero. There is also a false vacuum in which the value of $\sigma_1$ is
larger than that of the true vacuum. The term $\epsilon$ denotes the difference in energy 
densities of the false and true vacuum and it is positive in this case.

A topological soliton solution such as a cosmic string or magnetic monopole has a vanishing
field strength within its core due to continuity requirements. Consequently, the field strength
must increase from zero and approach the vacuum value asymptotically. Now if a monopole
is present in the false vacuum of figure \ref{potcur}, the core of this monopole contains a 
region in which the field strength corresponds to the true vacuum. Following the discussion
of \cite{Steinhardt:1981ec}, we assume that this region is a spherical bubble of radius
$R_b$ which has a thin boundary. This is referred to as the ``thin-wall" approximation. 
Under this approximation, the monopole consists of three regions. The inner-most region is
a spherical region which corresponds to the true vacuum. This is followed by a ``thin-wall" 
in which the field strength grows rapidly to value corresponding to the metastable vacuum. 
Outside this wall is a region of false vacuum in which the energy density is higher than that
of the true vacuum.

\begin{figure}[!htp]
\begin{center}
\includegraphics[width=0.8\textwidth]{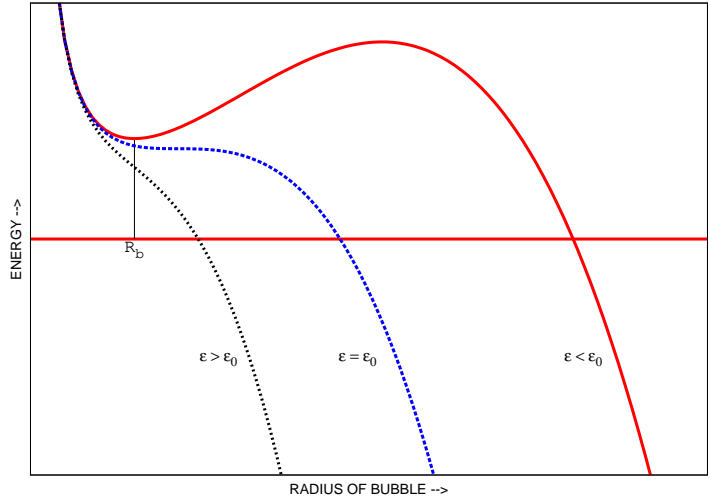}
\caption{The energy of a bubble of radius $R$. For small enough $\epsilon$, there is a 
minimum of $E(R)$ corresponding to a stable monopole solution with finite bubble radius. 
If $\epsilon$ becomes too large, there is no minimum and $R_b \rightarrow \infty$. This
corresponds to an expansion of the core resulting in an unstable monopole configuration.}
\label{bubbleenergy}
\end{center}
\end{figure}

A question which naturally arises is whether such a bubble of true vacuum can expand and occupy
the region of false vacuum outside the core of the monopole. If the bubble expands from a 
radius $R$ to $R + dR$, the change in energy to first order in $dR$ is $-4\pi \epsilon R^2dR$. 
The negative sign signifies that energy can be lost through radial expansion. Thus, the energy
of the bubble is proportional to $ -\epsilon R^3$. However, an increase in radius of the bubble
leads to an increase in surface area of the surrounding wall which has a positive energy density
$\sigma$. Hence, there is a term in the total bubble energy proportional to $\sigma R^2$. There is also 
a gauge field present in a topological soliton and its energy varies as $C/R$ for some constant
C. Thus, the total energy of the bubble is given by
\begin{equation}
E(R) = -\epsilon \frac{4\pi}{3} R^3 + 4\pi \sigma R^2 + \frac{C}{R}.
\label{buben}
\end{equation}
This energy as a function of bubble radius $R$ is plotted in figure \ref{bubbleenergy}. The profile
$E(R)$ has a local minimum for small values of $\epsilon$. The value of $R$ for this minimum is
denoted $R_b$ and this corresponds to a stable monopole solution. As $\epsilon$ increases, the 
curvature of the local minimum becomes less positive and at a critical value $\epsilon_0$, it 
becomes zero. If $\epsilon$ is increased beyond $\epsilon_0$, there is no local minimum and 
$R_b$ tends to infinity. This indicates an unstable monopole configuration whose core contains
a region of true vacuum which is expanding into the surrounding false vacuum.

In the supersymmetry breaking model we will consider, it turns out that value of $\epsilon$
is governed by a parameter $\tilde{M}$. Increasing $\tilde{M}$ increases both the 
VEV of $\sigma_1$ and also the energy of the false vacuum (see figure \ref{potcur}). Thus,
\begin{equation}
\epsilon \sim \tilde{M}^4.
\end{equation}
As will be shown in a later section, increasing $M$ beyond a certain value creates an instability
in the monopole configuration which settles into the metastable vacuum.

We will be studying instability from two approaches. One is by simply looking for time-independent
monopole solutions. If a monopole becomes classically unstable in the way described above, it is
described by a time-dependent solution describing an expanding bubble of true vacuum. In this case, 
it is impossible to obtain a time-independent solution. Hence, whenever a time-independent solution
cannot be obtained, that monopole configuration is understood to be unstable for those values of
parameters. This will be discussed in greater detail in section \ref{stabilitynonsusy}. We also 
follow a semi-analytic approach to the study of stability and this will be discussed in 
section \ref{semannonsusy}.

\section{The model and its vacua}
\label{modelandansatz}

We study a model described in \cite{Bajc:2008vk} which enables metastable
supersymmetry breaking without the use of singlet fields. This requires the
use of two $SU(5)$ adjoints $\Sigma_{1}$ and $\Sigma_{2}$ whose vacuum
expectation values (VEVs) are given by
\begin{equation}
\vev{\Sigma_{i}} = \frac{\vev{\sigma_{i}}}{\sqrt{30}} \left(
\begin{array}{ccccc}
2\, & 0\, & 0 & 0 & 0 \\
0\, & 2\, & 0 & 0 & 0 \\
0\, & 0\, & 2 & 0 & 0 \\
0\, & 0\, & 0 & -3 & 0 \\
0\, & 0\, & 0 & 0 & -3
\end{array} \right)
\label{vev}
\end{equation}
where $\sigma_{i}\,(i = 1,2)$ are the standard model singlets. The
superpotential is
\begin{eqnarray}
W &=& Tr\left[\Sigma_{2}\left(\mu \Sigma_{1} + \lambda \Sigma_{1}^{2} +
\frac{\alpha_{1}}{M}\Sigma_{1}^{3} +
\frac{\alpha_{2}}{M}Tr(\Sigma_{1}^2)\Sigma_{1} \right) \right] \nonumber \\
&=& \frac{1}{30M} \, \sigma_{2}\left( 30M\mu - \sqrt{30}M\lambda\sigma_{1}
+ (7\alpha_{1} + 30\alpha_{2})\sigma_{1}^{2}\right)\sigma_{1}
\label{superpot}
\end{eqnarray}
from which the scalar potential can be written as
\begin{eqnarray}
V &=& \left( \mu \sigma_{1} - \frac{\lambda \sigma_{1}^2}{\sqrt{30}} +
\frac{7\alpha_{1} \sigma_{1}^3}{30M} + \frac{\alpha_{2}\sigma_{1}^3}{M}
\right)^2 \nonumber \\
&+& \frac{\sigma_{2}^2}{900 M^2}\left( 30M\mu - 2\sqrt{30}M\lambda\sigma_{1}
+ 3(7\alpha_{1} + 30\alpha_{2})\sigma_{1}^2\right)^2.
\label{scalarpot}
\end{eqnarray}
There are no D-Term contributions to the scalar potential since the VEVs
of $\Sigma_i$ are diagonal. Defining $v_{1} = \vev{\sigma_{1}}$ and
$v_{2} = \vev{\sigma_{2}}$,
$SU(5)$ is broken at $v_2$ while supersymmetry is broken at $v_1$. The VEV
$v_1$ can be obtained from minimizing (\ref{scalarpot}) but $v_2$ is undetermined
at tree order. This flat direction is lifted by nontrivial 1-loop corrections to
the scalar potential which have the form $V \approx  m_{\sigma_2}^2(\sigma_2 - v_2)^2$.
Adding a term like this on the right hand side of (\ref{scalarpot}) leaves the value
of $v_1$ unchanged while $\vev{\sigma_2}$ gets fixed at $v_2$.
The units of $v_2$ are the same as those of $\sigma_2$ and can be taken to be $100\mu$
to simplify the forthcoming discussion.
We thus stabilize $v_2$ in 
the ensuing numerical simulations 
by adding $(\sigma_2 - v_2)^2$ to the scalar potential (\ref{scalarpot}) and hence 
fixing $\sigma_2$ to $v_2$ (in units of $100\mu$). An example of the scalar potential
with $v_2$ and the couplings fixed is shown in figure \ref{potcur}.

\subsection{Supersymmetry breaking vacua}
\label{susybreakingvacua}

The model we are considering has no supersymmetric minima unless $v_2 = 0$. When $SU(5)$ breaks
with $v_2 \neq 0$, there are two supersymmetry breaking vacua which we shall denote $\ket{V_1}$ and
$\ket{V_2}$. The value of $v_1$ for these minima is given by
\begin{equation}
v_1 = \frac{1}{3(7\alpha_1 + 30\alpha_2)} \left[ \sqrt{30}M\lambda \pm
\sqrt{30}\sqrt{M^2\lambda^2 - 21M\mu\alpha_1 - 90M\mu\alpha_2} \right].
\label{v1vevs}
\end{equation}
The couplings in this model are chosen in conformity with the requirements discussed
in \cite{Bajc:2008vk}. The variables $\mu$ and $M$ in the above equation have
dimensions of mass whereas $\lambda$, $\alpha_1$, and $\alpha_2$ are 
dimensionless. We define the dimensionless variable $\tilde{M}$ through the relation
$M = \tilde{M}\mu$, and choose 
$\alpha_1 = \alpha_2 = 0.1$ and
$\lambda = 0.5$.  The resulting expression for $v_1$ becomes
\begin{equation}
v_1 = \left[ \frac{1}{11.1} \left[ \frac{\sqrt{30}\tilde{M}}{2} \pm
\sqrt{30}\sqrt{\frac{\tilde{M}^2}{4} - 11.1\tilde{M}}\, \right] \right] \mu.
\end{equation}
The value of the dimensionless variable $\tilde{M}$ is varied later in the paper to study its effect on 
stability of vacua containing monopoles.
The $\pm$ signs in the above equation correspond to two distinct vacua, one near the origin due to almost
exact cancellation and one far from the origin. The vacuum which is far from the origin is
referred to as $\ket{V_1}$. Its properties for $\tilde{M} \approx 1000$ are described below:
\begin{equation}
\ket{V_1}: v_1 \approx 10^{2}\mu ~~~ \langle V_1\,|V|\,V_1 \rangle \approx 10^{8}\mu^4.
\label{vacuum1}
\end{equation}
In contrast, the state $\ket{V_2}$ has the following properties:
\begin{equation}
\ket{V_2}: v_1 \approx \mu ~~~ \langle V_2\,|V|\,V_2 \rangle \approx \mu^4.
\label{vacuum2}
\end{equation}
Both these minima have non-zero energy but the energy of $\ket{V_2}$ is much smaller
than that of $\ket{V_1}$. The state $\ket{V_2}$ is thus the true vacuum for this model
whereas $\ket{V_1}$ is metastable.
When monopole configurations can exist in the above vacua, they shall be denoted
$\ket{V_{1}^{monopole}}$ and $\ket{V_{2}^{monopole}}$ respectively. 

For a given value
of $v_2$, the configuration $\ket{V_{1}^{monopole}}$ which asymptotes to
$\ket{V_1}$ at infinity must pass through the field value $v_1 \approx \mu$
corresponding to $\ket{V_2}$ near the core of the monopole. This ``seeding'' of
the true vacuum which occurs for $\ket{V_{1}^{monopole}}$ inside its core can
render it unstable as will be shown in section \ref{stabilitynonsusy}. 
On the other hand, such a seeding does not take place for $\ket{V_{2}^{monopole}}$
and it is thus expected to remain stable. This expectation is confirmed by our
numerical results which will be discussed in section \ref{stabilitynonsusy}. 

The phenomenological motivations behind the choice of the superpotential given in
equation [\ref{superpot}] is described in \cite{Bajc:2008vk}. The two terms 
involving $\alpha_1$ and $\alpha_2$ are non-renormalizable terms which vanish
in the limit $M \rightarrow \infty$. The desired SUSY breaking vacuum of this 
model turns out to be $\ket{V_2}$ and this state survives the renormalizability limit.
The other vacuum $\ket{V_1}$ gets pushed further away from the origin in field
space as $M$ increases and eventually becomes tachyonic for large enough $M$. 

\subsection{Supersymmetric vacua}
\label{susyvacua}

This model contains two supersymmetric vacua when $v_2 = 0$. When this is the case,
the value of $v_1$ for these minima is given by:
\begin{equation}
v_1 = \frac{1}{2(7\alpha_1 + 30\alpha_2)} \left[ \sqrt{30}M\lambda \pm
\sqrt{30}\sqrt{M^2\lambda^2 - 28M\mu\alpha_1 - 120M\mu\alpha_2} \right].
\label{v1vevssusy}
\end{equation}
Once again, the $\pm$ signs yield one vacuum close to the origin and one far from it.
The vacuum further from the origin is denoted as $\ket{V_{1(SUSY)}}$ for which the
value of $v_1$ is written as $v{_1}^{+}$. The plus sign in the superscript signifies
that the plus sign in equation \ref{v1vevssusy} has been used. Likewise, the
vacuum near the origin is referred to as $\ket{V_{2(SUSY)}}$ for which $v_1$ is
given by $v_{1}^{-}$. Both these states have
exactly zero energy, but since $v_{1}^{-} < v_{1}^{+}$, a monopole configuration which
asymptotes into $v_{1}^{+}$ corresponding to $\ket{V_{1(SUSY)}}$
must pass through the field value $v_{1}^{-}$ corresponding to $\ket{V_{2(SUSY)}}$ near
the origin. Thus, a ``seeding'' effect similar that of $\ket{V_{1}^{monopole}}$ takes place
when a monopole is present in $\ket{V_{2(SUSY)}}$, rendering it unstable. This will be
discussed in greater detail in section \ref{stabilitysusy}.

\section{The monopole ansatz and equations of motion}
\label{eom}

We shall set up monopole configurations in the fields $\Sigma_1$ and $\Sigma_2$.
The Lagrangian for the system can be expressed as
\begin{equation}
L = -\frac{1}{4}F_{\mu\nu}^{a}F^{a\mu\nu} + \frac{1}{2}(D_{\mu}\Sigma_1)^2
+ \frac{1}{2}(D_{\mu}\Sigma_2)^2 - V(\Sigma_1,\Sigma_2)
\label{lag}
\end{equation}
where $a = 1,...,24$ and $F_{\mu\nu}^a$ are the gauge field strengths.
The covariant derivative is expanded as
\begin{equation}
D_{\mu}\Sigma_{i}^{a} = \partial_{\mu}\Sigma_{i}^{a} - ie[F_{\mu},\Sigma_{i}]^a
\end{equation}
with $i = 1,2$ and the adjoints are expressed in terms of the $SU(5)$ generators
$T^a$ as $\Sigma_{i} = \sigma_{i}^a T^a$. We choose a spherically symmetric
ansatz for the adjoints and the gauge field:
\begin{equation}
\sigma_{1}^a = \frac{r^a}{er^2}G(r),~~~\sigma_{2}^a = \frac{r^a}{er^2}H(r)
\label{sigmaans}
\end{equation}
\begin{equation}
A_{n}^{a} = \epsilon_{amn}\frac{r^m}{er^2}[1 - K(r)],~~~A_{0}^{a} = 0.
\label{aans}
\end{equation}
To setup the $SU(5)$ monopole, we take the following embedding of $SU(2)$ in
$SU(5)$:
\begin{equation}
\left(
\begin{array}{cccc}
0 & \, & \, & \, \\
\, & 0\, & \, & \,  \\
\, & \, & \tau_a & \,  \\
\, & \, & \, & 0  \\
\end{array} \right).
\end{equation}
Here, $\tau_a = \frac{1}{2}\sigma_a \,(a = 1,2,3)$, and $\sigma_a$ are the Pauli
Sigma matrices. The $2 \times 2$ matrices $\tau_a$ satisfy
$[\tau_i,\tau_j] = i\epsilon_{ijk}\tau_k$. The $SU(5)$ adjoints $\Sigma_1$ and
$\Sigma_2$ given in equation $\ref{vev}$ are aligned along the generator
$\frac{1}{\sqrt{30}}diag(2,2,2,-3,-3)$ in isospace. In order to obtain the
non-trivial topology corresponding to a monopole configuration, we perform a unitary
transformation on the fields $\Sigma_1$ and $\Sigma_2$ and also the gauge field $A$ to
obtain (see \cite{Meckes:2002gx})
\begin{equation}
\Sigma_{1} = \frac{1}{\sqrt{30}} \sum_{a=1}^{3} \left(
\begin{array}{cccc}
2\, & \, & \, & \,  \\
\, & 2\, & \, & \,  \\
\, & \, & -\frac{1}{2}I_2 + \frac{25}{er^2}G(5r)\tau_a r^a & \,\\
\, & \, & \, & -3
\end{array} \right)
\end{equation}
\begin{equation}
\Sigma_{2} = \frac{1}{\sqrt{30}} \sum_{a=1}^{3} \left(
\begin{array}{cccc}
2\, & \, & \, & \,  \\
\, & 2\, & \, & \,  \\
\, & \, & -\frac{1}{2}I_2 + \frac{25}{er^2} H(5r)\tau_a r^a & \,\\
\, & \, & \, & -3
\end{array} \right)
\end{equation}
\begin{equation}
A_n = \sum_{i,j=1}^{3} \left(
\begin{array}{cccc}
0\, & \, & \, & \,  \\
\, & 0\, & \, & \,  \\
\, & \, & \frac{25}{er^2}[1- K(5r)]\epsilon_{ijn}\tau_i r^j & \,\\
\, & \, & \, & 0
\end{array} \right)
\end{equation}
where $I_2$ is the $2 \times 2$ identity matrix. Using the above results in the
lagrangian (\ref{lag}), the Euler-Lagrange equations of motion become second-order
differential equations in $G(r)$, $H(r)$ and $K(r)$. Taking $e=1$, they are given by
\begin{equation}
r^2K^{\prime\prime} = K(K^2 - 1) + K(G^2 + H^2)
\label{keqn}
\end{equation}
\begin{equation}
r^2G^{\prime\prime} = 2GK^2 + r^4\frac{\partial}{\partial G}\left(V(G,H)\right)
\label{geqn}
\end{equation}
\begin{equation}
r^2H^{\prime\prime} = 2HK^2 +  r^4\frac{\partial}{\partial H}\left(V(G,H)\right)
\label{heqn}
\end{equation}
in which the function $V(G,H)$ is obtained by substituting $\sigma_1 =
G/r$ and $\sigma_2 = H/r$ in the scalar potential (\ref{scalarpot}). These equations
of motion are ordinary differential equations involving only $r$ as a result of our
spherically symmetric ansatz functions given in (\ref{sigmaans}) and (\ref{aans}).
The function $G(r)/r$ as $r \to \infty$ tends to a constant value. This means that
the function $G(r)$ is proportional to $r$ and it therefore diverges at infinity.
This is also true for the function $H(r)$. For the purpose
of our numerical solutions for the monopole, we express equations (\ref{keqn} -
\ref{heqn}) in terms of $\sigma_1$ and $\sigma_2$. The result is
\begin{equation}
\frac{d^2K}{dr^2} - \frac{1}{r^2}K(K^2-1) - K(\sigma_1^2 + \sigma_2^2) = 0
\label{keom}
\end{equation}
\begin{equation}
\frac{d^2\sigma_1}{d\sigma_1^2} + \frac{2}{r}\frac{d\sigma_1}{dr}
- \frac{2}{r^2}\sigma_1K^2 - \left( \frac{\partial \,V}{\partial \sigma_1} \right) = 0
\label{soneeom}
\end{equation}
\begin{equation}
\frac{d^2\sigma_2}{d\sigma_1^2} + \frac{2}{r}\frac{d\sigma_2}{dr}
- \frac{2}{r^2}\sigma_2K^2 - \left( \frac{\partial \,V}{\partial \sigma_2} \right) = 0.
\label{stwoeom}
\end{equation}

Equations (\ref{keom} - \ref{stwoeom}) can be solved numerically using relaxation
techniques after rescaling $\sigma_1$, $\sigma_2$ and the variable $r$ by the VEV $v_1$.
A monopole configuration necessarily implies $\sigma_1 = \sigma_2 = 0$ at
$r = 0$. From equation (\ref{keom}), the function $K(r) \to 1$ as $r \to 0$. As
$r \to \infty$, the function $K(r) \to 0$ while the values of $\sigma_1$ and $\sigma_2$
are determined by solving the following set of simultaneous polynomial equations:
\begin{equation}
\left( \frac{\partial \,V}{\partial \sigma_1} \right) = 0
\label{boundone}
\end{equation}
\begin{equation}
\left( \frac{\partial \,V}{\partial \sigma_2} \right) = 0.
\label{boundtwo}
\end{equation}
The initial guess for the solution is chosen to meet the boundary conditions discussed
above. The value of $M$ is varied to study its effect on the stability of
the vacua containing monopoles. For a given set of couplings, there are two solutions
to equations (\ref{boundone}) and (\ref{boundtwo}) corresponding to the two translation
invariant vacua described by (\ref{vacuum1}) and (\ref{vacuum2}).

\section{Stability of the monopole configurations in non-supersymmetric vacua}
\label{nonsusyg}

\subsection{Numerical study}
\label{stabilitynonsusy}

The numerical methods we use are relaxation techniques implemented by discretizing the
domain over which the solution is required. By dividing the interval into a sufficient 
number of points, we convert the differential equations into a set of simultaneous
polynomial equations. The initial guess is chosen in conformity with the boundary conditions
discussed at the end of section \ref{eom}. 

\begin{figure}[!htp]
\begin{center}
\includegraphics[width=0.8\textwidth]{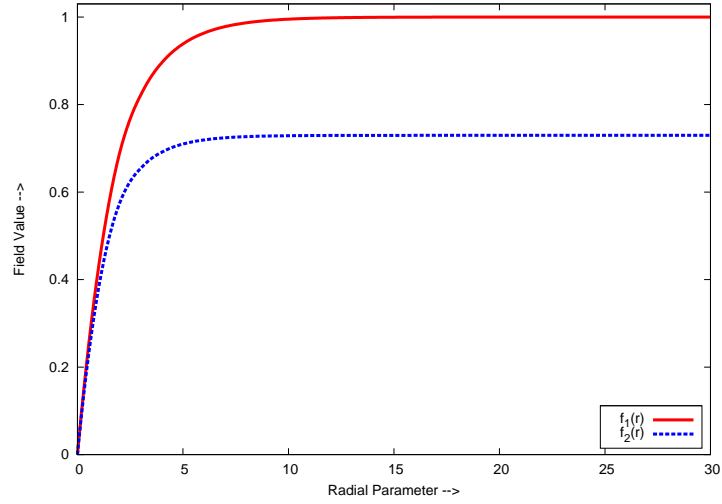}
\caption{The functions $f_1(r) = \sigma_1(r)/v_1$ and $f_2(r)= \sigma_2(r)/v_1$ for $\ket{V_1^{monopole}}$
with $\tilde{M} = 1400$ and $v_2 = 500\mu$. The value of $v_1$ is $685.3\mu$.}
\label{s12}
\end{center}
\end{figure}

\begin{figure}[!htp]
\begin{center}
\includegraphics[width=0.8\textwidth]{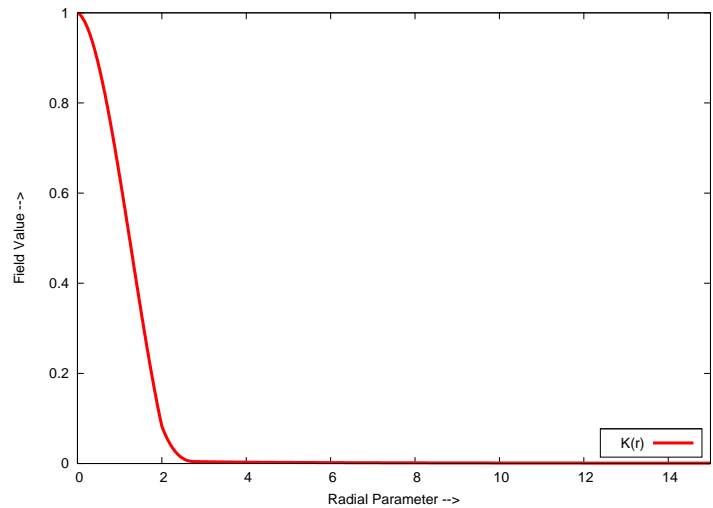}
\caption{The function $K(r)$ for $\ket{V_1^{monopole}}$ with $v_2 = 500\mu$ and $\tilde{M} = 1400$.}
\label{kplot}
\end{center}
\end{figure}

We denote by $\ket{V_1^{monopole}}$ the monopole solution which settles into
$\ket{V_1}$ at infinity. Likewise, $\ket{V_2^{monopole}}$ denotes the monopole configuration
which reaches $\ket{V_2}$ asymptotically. We have studied the availability of both
$\ket{V_1^{monopole}}$ and $\ket{V_2^{monopole}}$ by first choosing a value of $v_2$
and then varying $\tilde{M}$. An example of $\ket{V_1^{monopole}}$ when $\tilde{M}=1400$ and 
$\sigma_2 = 500\mu$ is shown in figures \ref{s12} and \ref{kplot}. In this case, the value of $v_1$ is 
$685.3\mu$ and this is independent of $\sigma_2$. A similar solution exists for $\ket{V_2^{monopole}}$ 
for which the value of $v_1$ is $5.5\mu$.

\begin{figure}[!htp]
\begin{center}
\includegraphics[width=0.8\textwidth]{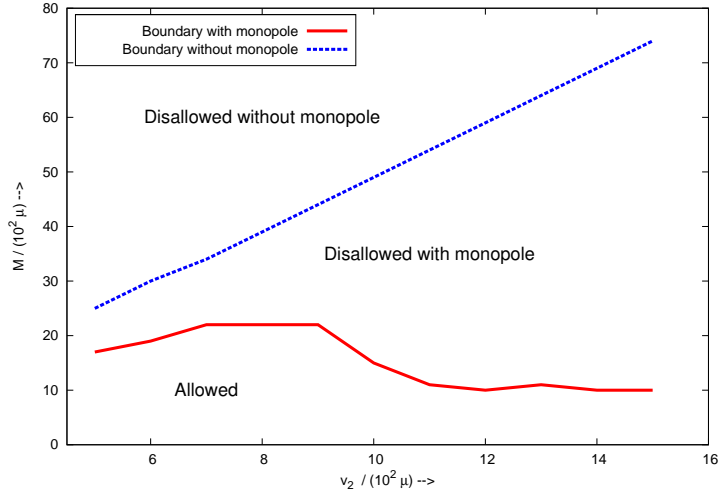}
\caption{The allowed and disallowed regions for $\ket{V_1^{monopole}}$. The boundary
between allowed and disallowed regions without a monopole present is defined by equation
(\ref{bound}). When a monopole is is present in $\ket{V_1}$, a large region of the
parameter space becomes disallowed since $\ket{V_1^{monopole}}$ is no longer available
for this region.}
\label{regions}
\end{center}
\end{figure}

As discussed in section \ref{modelandansatz}, there is a ``seeding'' of the true vacuum within the
core of $\ket{V_1^{monopole}}$. In the above example, the field value
$\sigma_1 = v_1 = 5.5\mu$ corresponds to the true vacuum 
denoted by $\ket{V_2}$. The monopole configuration for $\sigma_1$ shown in figure
\ref{s12} must rise from zero and pass through $\sigma_1 = 5.5\mu$ before reaching its
asymptotic value of $685.3\mu$. Furthermore, the energy of the local minimum at
$\sigma_1 = 685.3\mu$ increases as $\tilde{M}$ increases. Therefore, for sufficiently large values
of $\tilde{M}$, we expect an instability in the configuration $\ket{V_1^{monopole}}$. This
expectation is confirmed by the results of our numerical simulations which are
summarized in figure \ref{regions}. There is a large region in the parameter space of
$v_2$ and $\tilde{M}$ for which a numerical solution for $\ket{V_1^{monopole}}$ cannot be obtained.

The value of $\tilde{M}$ cannot be increased to arbitrarily high values. There is a limiting value
of $\tilde{M}$ beyond which the metastable vacuum $\ket{V_1}$ becomes tachyonic. The condition
which must be satisfied for no tachyonic states is \cite{Bajc:2008vk}
\begin{equation}
2\left| \frac{v_2}{v_1} \right|^2 \left| \frac{F}{v_1^2} - \left( \frac{7}{30}\alpha_1
+ \alpha_2 \right)\frac{v_1}{M} \right| \geq \left| \frac{F}{v_1^2} \right|
\label{bound}
\end{equation}
in which $F$ is defined as
\begin{equation}
F = v_1^2 \left[ \frac{\lambda}{\sqrt{30}} - \frac{2}{M}\left(\frac{7}{30}\alpha_1
+ \alpha_2 \right)v_1 \right].
\end{equation}
This condition puts an upper bound on possible values of $\tilde{M}$ as shown in figure
\ref{regions}. However, when a monopole is present in $\ket{V_1}$, this upper bound
becomes significantly lowered.

The fact that the numerical solution for $\ket{V_1^{monopole}}$ cannot be obtained
for a finite region of the parameter space is indicative of an inherent instability
in such a configuration. This instability will be studied from a semi-analytic approach
in the next subsection. The configuration $\ket{V_2^{monopole}}$ can be obtained for
all of the parameter values which were considered. Such a configuration has no lower energy
state to decay into.

\subsection{Semi-Analytic approach}
\label{semannonsusy}

The numerical solutions obtained in the previous subsection are all time-independent. We can
restore time-dependence in the equations of motion and study the stability of the solutions
without actually solving the time-dependent equations of motion. Following the approach
discussed in \cite{Yajnik:1986wq}, we restore time-dependence in equation (\ref{soneeom}):
\begin{equation}
-\frac{d^2\sigma_1}{dt^2} + \frac{d^2\sigma_1}{d\sigma_1^2} + \frac{2}{r}\frac{d\sigma_1}{dr}
- \frac{2}{r^2}\sigma_1K^2 - \left( \frac{\partial \,V}{\partial \sigma_1} \right) = 0.
\label{soneteom}
\end{equation}
When studying the possibility of  $\ket{V_1^{monopole}}$ decaying into
$\ket{V_2^{monopole}}$, it is sufficient to study the time-dependence of $\sigma_1$ alone.
This is because the vacuum value of $\sigma_2$ is equal to $v_2$ for both these states.
We thus treat $\sigma_2$ as a time-independent background field denoted by $\tilde{\sigma}_2(r)$
in the following discussion. The time-dependence of $\sigma_1(r,t)$ is decomposed as follows:
\begin{equation}
\sigma_1(r,t) = \tilde{\sigma}_1(r) + p(r)e^{i\omega t}.
\label{timedeps1}
\end{equation}
The function $p(r) << \tilde{\sigma}_1(r)$ and $\tilde{\sigma}_1(r)$ is the
time-independent solution to
equation (\ref{soneeom}). Substituting equation (\ref{timedeps1}) in (\ref{soneteom}) and
linearizing in $p$, we obtain
\begin{equation}
\omega^2 p = -\left[ \frac{d^2}{dr^2} + \frac{2}{r}\frac{d}{dr} \right] p + \left[U(r)\right]p.
\label{eqsch}
\end{equation}
This equation has the form of a one-dimensional Schrodinger equation with a potential
\begin{eqnarray}
U(r)/10^4\mu^2 &=&  \beta \left( 147{\alpha_1}^2 + 2700{\alpha_2}^2 + 1260\alpha_1\alpha_2 \right) 
 \left( 18\tilde{\sigma}_2^2\tilde{\sigma}_1^2 + 5\tilde{\sigma}_1^4 \right) \nonumber \\
& & -\,20\beta \tilde{M}\tilde{\sigma}_1^2\left( 7\alpha_1  + 30\alpha_2 \right) 
   \left( {\sqrt{30}}\lambda\tilde{\sigma}_1
 - 0.18 \right) \nonumber \\ 
& & +\, 60\beta \tilde{M}^2\left( 2{\lambda }^2\tilde{\sigma}_2^2 + 3{\lambda }^2\tilde{\sigma}_1^2 
 - 0.03{\sqrt{30}}\lambda \tilde{\sigma}_1  + 0.0015 \right) \nonumber \\
& & +\, 36\beta \tilde{M}\tilde{\sigma}_2^2\left( 7\alpha_1  + 30\alpha_2 \right) \left( 
   {\sqrt{30}}\lambda\ \tilde{\sigma}_1  -
    0.05  \right) + \,\frac{2\tilde{K}^2}{r^2}
\label{equivpot}
\end{eqnarray}
where $\beta = 1/450 \tilde{M}^2$ and $\tilde{\sigma}_1$, $\tilde{\sigma}_2$, and $\tilde{K}$ are
time-independent dimensionless functions of $r$. 

\begin{figure}[!htp]
\begin{center}
\includegraphics[width=0.8\textwidth]{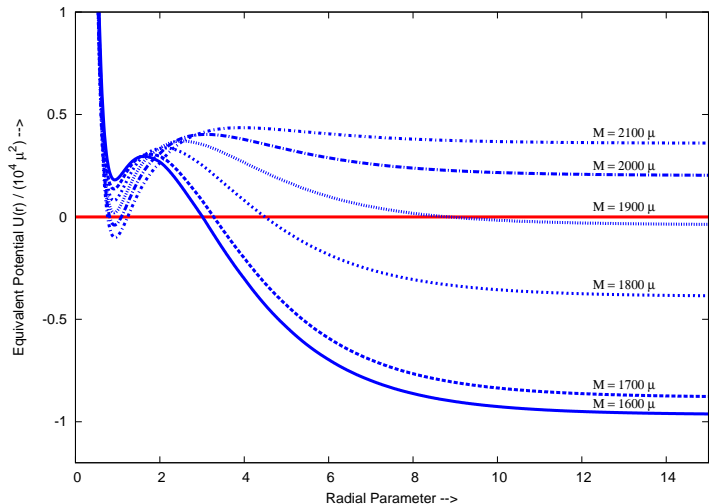}
\caption{The equivalent potential for $\ket{V_1^{monopole}}$ with $\sigma_2 = 500\mu$.
The effect of increasing $\tilde{M}$ can be seen to lower the energy of the local minimum
near the origin. For large enough $\tilde{M}$, this minimum has a negative energy and the
resulting bound state creates an instability in $\ket{V_1^{monopole}}$.}
\label{ep1}
\end{center}
\end{figure}

The stability of $\ket{V_1^{monopole}}$ depends on whether or not
the frequencies of oscillation $\omega$ in equation (\ref{timedeps1}) are imaginary. Real
frequencies result in stable solutions whereas imaginary frequencies indicate instability.
Looking for imaginary modes of oscillation or negative values of $\omega^2$ is equivalent to
looking for negative energy bound states of the potential given in equation (\ref{equivpot}).
This potential is plotted for different values of $\tilde{M}$ with $\sigma_2 = 5$ for
$\ket{V_1^{monopole}}$ in figure \ref{ep1}. Notice that there is a local minimum
near the origin, and that the energy of this minimum reduces as $\tilde{M}$
increases.  What this means is that
$\ket{V_1^{monopole}}$ is only stable for small enough values of $\tilde{M}$. When this is so, there is
no negative energy bound state possible. When $\tilde{M}$ is large enough, a negative energy
bound state is possible and the configuration $\ket{V_1^{monopole}}$
is no longer stable.

\section{Stability of the monopole configurations in supersymmetric vacua}
\label{stabilitysusy}

For the case of the supersymmetric vacua, the monopole configuration settling into
$\ket{V_{1(SUSY)}}$ is denoted $\ket{V_{1(SUSY)}^{monopole}}$, and likewise for
$\ket{V_{2(SUSY)}^{monopole}}$. It turns out that the numerical solutions for
$\ket{V_{1(SUSY)}^{monopole}}$ for values of $\tilde{M}$ larger than 200
can not be obtained. As discussed in the previous section, this failure to obtain
a numerical solution is indicative of an instability of the configuration. Thus, even 
a supersymmetric vacuum containing a monopole can become unstable against decay into 
another supersymmetric vacuum.

\begin{figure}[!htp]
\begin{center}
\includegraphics[width=0.8\textwidth]{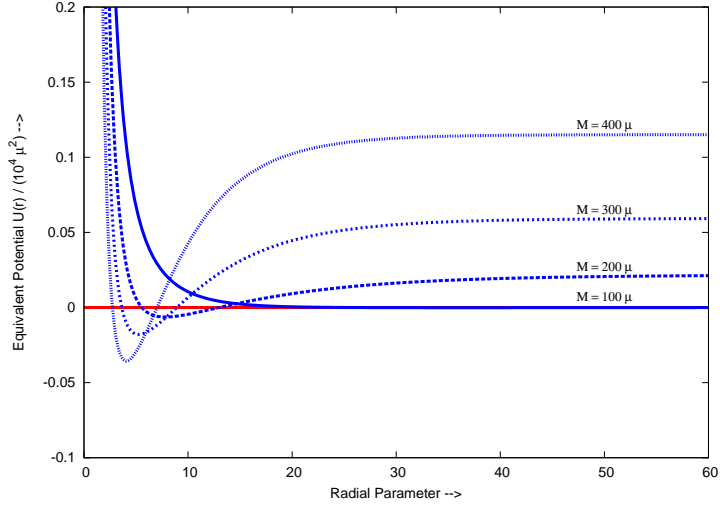}
\caption{The equivalent potential for $\ket{V_{1(SUSY)}^{monopole}}$.
For values of $\tilde{M} \gtrsim 200$, the negative energy bound state of the potential
results in an instability in $\ket{V_{1(SUSY)}^{monopole}}$.}
\label{ep2}
\end{center}
\end{figure}

With regards to the stability analyzed from the semi-analytic approach of the previous
section, the equivalent Schrodinger potential for $\ket{V_{1(SUSY)}^{monopole}}$ can be derived
in the same manner as was done for $\ket{V_1^{monopole}}$. It is given by
\begin{eqnarray}
U(r)/10^4\mu^2 &=& \gamma \tilde{\sigma}_1^4\left( 147 \alpha_1^2
 + 2700\alpha_2^2 + 1260\alpha_1\alpha_2 \right)  \nonumber \\
& & + \,36\gamma\alpha_2 \tilde{M}^2\,\left( \tilde{\sigma}_1^2{\lambda }^2 - 
          0.01{\sqrt{30}}\,\tilde{\sigma}_1\lambda + 0.0005 \right)  \nonumber \\
& & +\, 4\tilde{M}\gamma\tilde{\sigma}_1^2 \left( 7\alpha_1 + 30\alpha_2 \right)
    \left(0.18 -  {\sqrt{30}}\,\lambda{\tilde{\sigma}_1} \right) + \,\frac{2}{r^2}\tilde{K}^2 
\end{eqnarray}
where $\gamma = 1/90 \tilde{M}^2$. This potential
is plotted in figure \ref{ep2} and once again, negative energy bound states
indicate instability. In this case, values of $\tilde{M}$ lying near 200 represent the boundary
between stable and unstable configurations.

\section{Implications for cosmology}
\label{impl}

The early universe is characterized by a monotonic reduction in temperature. From 
calculations of thermal effective potentials 
\cite{Weinberg:1972ws,Kirzhnits:1972ut,Dolan:1973qd}, we expect a leading order
temperature correction of the form $AT^2\sigma_1^2$ with $A > 0$ to the effective
potential of $\sigma_1$. If we absorb this effect in the definition of $\mu_{eff}$, 
we have
\begin{equation}
\mu_{eff}^2 = \mu^2 + AT^2.
\end{equation}
If we understand
the rescaled quantity $\tilde{M}$ to be expressed in terms of $\mu_{eff}$, we see that
a change in $\tilde{M}$ is equivalent to a change in $T$. Specifically, we have 
$M/\mu_{eff} = \tilde{M}$, which shows how an increase in temperature is equivalent
to a reduction of $\tilde{M}$. 

In this way, the increase of $\tilde{M}$ which results in an instability in 
$\ket{V_1^{monopole}}$ can happen naturally in the early universe with $M$ fixed
and the temperature $T$ decreasing. Referring to figure \ref{regions}, we see
that for a fixed value of $v_2 = \vev{\sigma_2}$, we may start with a high 
temperature period in the early universe which supports the existence of
$\ket{V_1^{monopole}}$. This high temperature is equivalent to low values
of $\tilde{M}$. As the temperature drops, the value of $\tilde{M}$
effectively increases and at a critical temperature $T_1$, 
$\ket{V_1^{monopole}}$ becomes unstable. Below this temperature, the solution
$\ket{V_1^{monopole}}$ is no longer available and this region is referred to as
``disallowed with monopole" in figure \ref{regions}. As the temperature continues to
drop, $\tilde{M}$ continues to rise until the state $\ket{V_1}$ becomes 
tachyonic at a temperature $T_2 < T_1$. For temperatures below $T_2$, the 
translation invariant vacuum $\ket{V_1}$ is unavailable and this is referred 
to as ``disallowed without monopole'' in figure \ref{regions}.

There is hence an intermediate range of temperatures $T_2 < T < T_1$ within 
which $\ket{V_1}$ is available but $\ket{V_1^{monopole}}$ is disallowed. When 
the universe passes through this temperature range, it is possible that causally 
connected local 
domains settle in the metastable state $\ket{V_1}$ instead of the true vacuum 
$\ket{V_2}$. Without
the presence of monopoles in $\ket{V_1}$, the transition to $\ket{V_2}$ can
only take place via quantum tunneling. A first-order phase transition of this
type can result in an inhomogeneous universe \cite{Guth:1982pn, Hawking:1982ga}.

In contrast, if monopoles had formed at higher temperatures in $\ket{V_1}$, the 
resulting states $\ket{V_1^{monopole}}$ would become unstable within this intermediate
temperature range. This would result in a classical roll-over to $\ket{V_2}$
and prevent an otherwise inhomogeneous expansion. In this way, the presence
of monopoles within $\ket{V_1}$ ensures a graceful exit from this false vacuum. 
The universe arising out of the type
of SUSY breaking model discussed here is hence homogeneous like our present day universe.

\section{Concluding Remarks}
\label{conc}

The results of this paper share a common theme with the results of 
\cite{Kumar:2008jb}. In both cases, a metastable vacuum containing a topological 
defect is shown to become unstable against a classical roll-over to the true vacuum.
However, the exact details of this process depend on the specific model being considered. In 
\cite{Kumar:2008jb}, the model of gauge mediated supersymmetry breaking 
which we had studied permitted the existence of cosmic strings. These
string configurations were responsible for the seeding of true vacuum bubbles
within their cores. Furthermore, the metastable vacua which became unstable 
in the presence of the strings were the phenomenologically desired SUSY breaking
minima. The true vacuum was an undesirable minimum in which $SU(3)_c$ was broken.
In this way, the cosmic strings played a potentially disastrous role in the model.

The role of the topological soliton is quite the opposite and much more benign in 
the present case. Here, the $SU(5)$ grand unified group supports the existence of 
monopoles and as we have discussed, the metastable vacua are phenomenologically 
undesirable. There is hence the need for a graceful exit from the false vacuum and
this is where the monopoles come in. If the universe is trapped in a
false vacuum and decays by quantum tunneling, the resulting universe can be 
inhomogeneous.
The type of classical monopole instabilities discussed here ensure that the metastable
vacuum can become unstable and undergo a ``roll-over" to the global minimum. 
This effect hence provides a graceful exit
from a potentially inhomogeneous expansion.

In general, any model based on a gauge group which supports the existence of 
topological solitons can undergo the process described here. Whether or not the
metastable vacua containing solitons are desirable depends on the given model. 
Hence, the stability of a metastable vacuum against quantum mechanical tunneling
is not always a sufficient condition to ensure the viability of a model 
containing solitons. 

\section{Acknowledgements}
We thank Borut Bajc and P. Ramadevi for useful comments and suggestions regarding this work.

\bibliographystyle{elsarticle-num}

\end{document}